\documentclass[epj]{svjour}
\usepackage[]{graphicx,float,subfig}
\usepackage{amsmath}
\usepackage{amssymb}
\usepackage{bbm}

\usepackage{color}
\definecolor{g_blue}{rgb}{0.2472,0.24,0.6}
\definecolor{g_red}{rgb}{0.6, 0.24, 0.442893}
\definecolor{g_gold}{rgb}{0.6, 0.547014, 0.24}
\definecolor{g_green}{rgb}{0.24, 0.6, 0.33692}

\newcommand{\legendLn}{\raise 0.4mm \hbox{$ \centerdot \!\! \centerdot \!\! \centerdot \!\! \centerdot \!\! \centerdot \!\! \centerdot \!\! \centerdot$}}
\newcommand{\vc}[1]{\textbf{\emph{#1}}}

\begin{document}
\title{Metachronal waves in a chain of rowers with hydrodynamic interactions}
\author{C. Wollin\inst{1} \and H. Stark\inst{1}
}                     
\offprints{H. Stark}          
\institute{Institut f\"ur Theoretische Physik, Technische Universit\"at 
Berlin, Hardenbergstr. 36, D-10623 Berlin Germany} 
%
%
\date{Received: date / Revised version: date}
%
\abstract{
Filaments on the surface of a microorganism such as Paramecium
or Ophalina beat highly synchronized and form so-called metachronal 
waves that travel along the surfaces. In order to
study under what principal conditions these waves form, we introduce a chain of
beads, called rowers, each periodically driven by an external force
on a straight line segment. To implement hydrodynamic interactions
between the beads, they are considered point-like. Two beads 
synchronize in antiphase or in phase depending on the positive or
negative curvature of their driving-force potential. Concentrating
on in-phase synchronizing rowers, we find that they display only 
transient synchronization in a bulk fluid. On the other hand,
metachronal waves with wavelengths of 7-10 rower distances emerge,
when we restrict the range of hydrodynamic interactions either artificially
to nearest neighbors or by the presence of a bounding surface 
as in any relevant biological system.
%
\PACS{
      {PACS-key}{discribing text of that key}   \and
      {PACS-key}{discribing text of that key}
     } 
} 
\maketitle
\section{Introduction}
For microorganisms with sizes between 1 to $1000\mu \mathrm{m}$ the 
surrounding fluid appears to be very viscous. They, therefore, need
a highly specialized lo\-co\-mo\-tion appa\-ra\-tus in order to propel
themselves \cite{Bray2001,Purcell77,Lauga09,downtonStark}.
Paramecium, \emph{e.g.}, uses a dense array of whip-like 
appendages on its surface, called cilia, that move fluid close to its surface
by beating periodically with a succession of power and recovery strokes 
\cite{machemerParamecium}.
The properties of cilia have been studied for decades \cite{Linck01}.
In particular, the coordinated beat cycles of a multitude of them, 
observable as traveling waves \cite{Brennen77}, has received a lot of 
attention recently
\cite{lagomarsinoEtAl,vilfanEtAl,Lenz06,niedermayerEtAl,kotarEtAl,Gueron97,gueronEtAl,joannyEtAl,Gauger09}.
The formation of traveling or so-called metachronal waves needs a high
degree of synchronization between the beating cilia. Many different types
of these waves called sym-, anti-, laeo- or dexioplectic are found on 
a variety of ciliated microorganisms, on certain types of comb jelly, and 
in the respiratory tract of mammals
\cite{blake_opalina_1975,opalinaSleigh,ctenophoreCiliaTamm,gillCiliaGibbons,tracheal_ephithelia_sleighEtAl,humRespCilia}.

Synchronization 
in large populations of interacting dynamic 
elements can be observed in many physical, chemical, biological, as well 
as social systems and a unifying description was formulated by Kuramoto 
\cite{kuramoto}. 
Examples range from micromechanic resonators used for the construction of 
highly sensitive mass-, spin-, and charge-mea\-su\-ring 
devices \cite{massSensing,spinDetection,electrometer},
over the self-organized beat cycle of single cilia, where the action of
dynein motors is coordinated by the curvature of the cilium 
\cite{riedelkruseflagellen},
the synchronized flashing of fireflies, 
the working of muscles and neurons, to the synchronization of the 
applause by an audience after the last bar of a classic concert 
\cite{syncr_Definitions,syncrRosenblumEtAl}.


In particular, in viscous fluids hydrodynamic coupling leads to the 
synchronization of rotating helical filaments such as bacterial
flagella \cite{Kim:2004,Reichert:2005}, rotating paddles
\cite{powersEtAl}, microfluidic rotors 
\cite{Grzybowski:2000,Lenz:2003,golestanian2010EtAl}, 
the pair of beating flagella in Chlamydomonas \cite{goldsteinEtAl}, 
or flagella of neighboring sperm cells 
\cite{Taylor51,gomperEtAl08,laugaEtAl09,Elfring10}. 
The most astonishing 
example for synchronization in viscous fluids are metachronal waves. 
A lot of works have been devoted towards understanding their formation.
They either use minimal models that abstract cilia as point-like objects
in order to identify essential features
\cite{lagomarsinoEtAl,vilfanEtAl,Lenz06,niedermayerEtAl,kotarEtAl}
or are based on a more accurate modeling of the beating cilia
\cite{Gueron97,gueronEtAl,joannyEtAl,Gauger09}.
From all these studies the common picture has evolved that hydrodynamic
interactions can synchronize beating cilia and thereby cause the
formation of metachronal waves.

This work uses the minimal model of Cosentino Lagomarsino \textit{et al.} 
\cite{lagomarsinoEtAl} that is based on a linear array of driven 
oscillators, so-called rowers, to clarify under what conditions
metachronal waves are able to form. First, we will demonstrate that the
long-range nature of hydrodynamic interactions impede the formation
of metachronal waves. Only when they become short-ranged close to
bounding surfaces do metachronal waves occur. Secondly, a stroke that
becomes faster during one half cycle leads to in-phase synchronization
of a pair of oscillators and thereby to metachronal waves in a linear array
of rowers with relatively long wavelengths. Clever recent experiments
of a pair of driven colloids have demonstrated antiphase 
synchronization when the stroke slows down during one half cycle
\cite{kotarEtAl}. It is our hope that these experiments can also 
demonstrate in-phase synchronization.

The paper is organized as follows. In Sec.\ \ref{sec_the_model} we introduce
our model including the potential for the force to drive the rowers. In
Sec.\ \ref{sec_sync_2_rowers} we investigate the synchronization of
two rowers by distinguishing between negative and positive curvature
in the driving-force potential. Then we study synchronization in a chain
of rowers in Sec. \ref{sec_sync_chain}. We first introduce an order parameter
for identifying metachronal waves, demonstrate their formation in a bulk
fluid when we artificially restrict hydrodynamic interactions to 
nearest-neighbor rowers, and show that a bounding surface has the same 
effect. We finish with a conclusion.

\section{The model} \label{sec_the_model}
In order to study basic features of metachronal waves, we abstract cilia as 
point-like beads following the successful works of Cosentino Lagomarsino \textit{et al.} 
\cite{lagomarsinoEtAl}, Niedermayer \textit{et al.} \cite{niedermayerEtAl}, 
and Vilfan and J\"ulicher \cite{vilfanEtAl}. We place the beads on a linear, 
periodic array and constrain their motions onto line segments, on which they
move back and forth. This model was first introduced by Cosentino Lagomarsino \textit{et
al.} and we adopt their convention where a bead is called rower
\cite{lagomarsinoEtAl}.

At low Reynolds number, the dynamics of the rowers 
is completely overdamped and the equations of motion for $N$ driven beads
read
\begin{equation}
  \dot{\vc{r}}_{m} = \sum_{n= 1}^{N} \vc{M}_{mn} (\vc{r}_{mn}) \vc{F}_{n} \,.
\label{dynamiceq}
\end{equation}
Here $\vc{r}_{m}$ is the position vector of bead $m$ and $\vc{F}_{m}$ the force
acting on it. The mobility matrices $ \vc{M}_{mn}$ depend on the difference
vector $\vc{r}_{mn} = \vc{r}_{m} - \vc{r}_{n}$. In an unbounded fluid with
viscosity $\eta$ and in the approximation of point-like particles, one uses
\begin{equation}
\vc{M}_{mn} ( \vc{r} ) =  \left\{
    \begin{array}{l l}
	\, \mu_0 \mathbbm{1} = \frac{1}{6 \pi \eta a} \mathbbm{1} & 
                                 \quad \text{if } n = m \\
        \, \vc{G}^{\text{Os}} = \frac{1}{8 \pi \eta } \frac{1}{r}
	    \left( \mathbbm{1} + \frac{ \vc{r} \otimes \vc{r}}{ r^{2} } \right)
                        & \quad \text{if } n \neq m \,, \\
    \end{array} \right.
\label{Oseen-Tensor}
\end{equation}
where $\mu_0 = 1/(6 \pi \eta a)$ is the Stokes mobility of a particle with
radius $a$, $\vc{G}^{\text{Os}}(\vc{r})$ the Oseen tensor, $\otimes$ means
dyadic product, and $r = |\vc{r}|$. In the course of this paper, we will also
study the chain of rowers close to an infinitely extended plane wall that 
bounds the fluid as depicted in Fig. \ref{chain_over_wall}. Then, the
mobilities of point particles are more complicated. We summarize them in 
Appendix \ref{app.wall}. In particular, the cross mobilities are given by 
the Blake tensor $\vc{G}^{\text{Bl}}(\vc{r})$. In the following, we employ
a rescaled version of the dynamic equations (\ref{dynamiceq}) by introducing
reduced quantities. We write the mobilities in units of $\mu_0$, the 
applied forces in units of a characteristic force value $F_0$, and lengths 
in units of $s$, where $2s$ is the length of a rower's line segment. Then time
is measured in units of $s/(\mu_0 F_0)$.

\begin{figure}
\centering
\subfloat{ \includegraphics[scale=0.8]{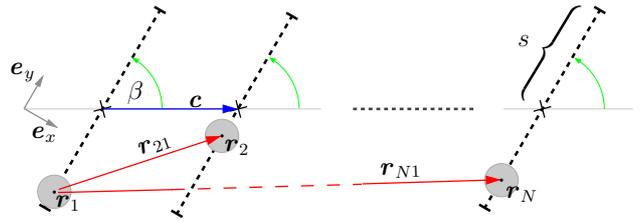} }
\caption{Geometry of the open linear chain of rowers that move back and 
forth on line segments (dashed lines) with length $2s$. The line segments
are oriented along the $y$ axis, their centers are separated by the
lattice vector $\vc{c}$, and the tilt angle between the segments and
$\vc{c}$ is $\beta$.
}
\label{geometry}
\end{figure}

Figure\ \ref{geometry} depicts the geometry of the linear chain of rowers 
separated by the lattice vector $\vc{c}$. The chain has open ends for which
we present most of our results. The effect of periodic boundary 
conditions is briefly discussed in Sec. \ref{subsec_unbound_fluid}.
While Cosentino Lagomarsino \textit{et al.} and Kotar \textit{et al.} let the rowers beat parallel 
to the lattice vector $\vc{c}$ \cite{lagomarsinoEtAl,kotarEtAl}, we 
introduce the tilt angle $\beta$ of line segments against the lattice vector 
$\vc{c}$ as global parameter.
The line segments are oriented along 
the $y$ axis and their centers are located at 
$m \vc{c} =  m c (\sin \beta,\cos \beta)$. Together with the asymmetry 
of the power and recovery stroke introduced below, this tilt breaks the 
left-right symmetry of the linear chain of rowers and metachronal waves can 
then propagate with a definite direction either to the left or to the right.


The driving forces acting on the rowers are always parallel to the line 
segments. Furthermore, we only consider the $y$ components of the 
hydrodynamic-interaction terms in Eq. (\ref{dynamiceq}) and neglect 
their $x$ components. This reduces the spatial degrees of freedom of each 
rower to one and its position is given by $ \vc{r}_{m} = m \vc{c} 
+ y_{m} \vc{e}_y $ (Fig. \ref{geometry}). In order to check this
approximation, we allowed excursions of the rowers perpendicular to their
line segments by introducing strong harmonic potentials along the 
$x$ direction. In simulations of this two-dimensional rower model,
we only found minor quantitative but no qualitative differences to the 
results presented below in various examples. We therefore decided
to work with the one-dimensional model, which needs much less simulation time.


In addition to the continous displacement variable $y_{m}$, we use the 
discrete ``geometric-switch'' variable $\sigma_{m} = \pm 1$ to describe the
rower's state. It was first introduced by Gueron and Levit-Gurevich 
\cite{gueronEtAl}.  It determines the direction of the driving force and 
hence if the rower moves parallel to $\vc{e}_y$ ($\sigma_{m} = 1$) or
antiparallel to $\vc{e}_y$ ($\sigma_{m} = -1$). The geometric switch changes 
sign when the rower reaches the displacement $s$ from the center of the
line segment. Therefore, the rower performs a periodic motion and can be
regarded as a driven oscillator.
For a driving force with constant magnitude $F_0$ and non-interacting rowers, 
the period of this oscillation is $\tau = 4 s / (\mu_0 F_{0}) $. In several
figures presented below we will refer time $t$ to the period $\tau$. For
later use we assign to each rower a phase variable that grows by $2 \pi$
during one cycle starting from the center of the line segment ($y_m = 0$):
\begin{equation}
\varphi_{m} \! = 2 \pi n_{m} + \frac{\pi}{ 2 } \sigma_{m} y_{m}
     + \begin{cases}
	  0 & \text{if } y_{m} \in [ \;\; 0 , \;\; 1 ) \, \wedge \, 
          \sigma_{m} \! = \;\: 1 \\
	\pi & \text{if } y_{m} \in [ \;\; 1 , - 1) \, \wedge \, 
          \sigma_{m} \! = \! - \! 1 \\
      2 \pi & \text{if } y_{m}  \in [ -1 , \;\: 0 ) \, \wedge \, 
           \sigma_{m} \! = \;\: 1 
        \end{cases}
\end{equation}
Here, $ n_{m} \in \mathbb{N}$ is increased by one after each completion 
of a cycle.

\begin{figure}
\centering
\subfloat{ \includegraphics[scale=0.9]{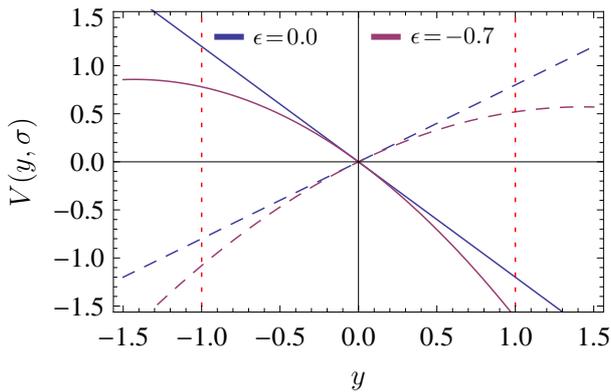} }
\caption{Potential of the driving force  for $\epsilon = 0$ and $\epsilon = -0.7$
and $\alpha=0.2$. The solid lines belong to $ \sigma = 1 $ and the dashed 
lines  to $ \sigma = -1$. For positive $\alpha$ the fast power stroke is
performed from left to right.
}
\label{potential}
\end{figure}

We define the reduced driving force on the $m$-th rower, 
$F_{m}( y_{m}, \sigma_{m} ) = - \partial V ( y_{m}, \sigma_{m}) / 
\partial y_m $, using the reduced potential
\begin{equation}
V ( y_{m}, \sigma_{m} ) = \left( -\sigma_{m} y_{m} + 
\frac{ \epsilon y_{m}^2 }{ 2 }  \right) ( \alpha \sigma_{m} + 1 ) + V_{P} \, .
\end{equation}
The linear term on the right-hand side gives a constant force with magnitude 
$F_0$ and the second term contributes a harmonic part to the potential with 
curvature $ \epsilon \! \in \! (-1, 1) $. The asymmetry parameter 
$\alpha \! \in \! ( -1, 1 )$ distinguishes between a fast power and a slow
recovery stroke.
Curvature $\epsilon$ and asymmetry $\alpha$ are the two important
parameters of the potential. Finally, we also introduce a harmonic potential
$V_{P}$ centered at the switching points ($y_m = \pm 1$). It acts only when 
the rower passes the switching point due to hydrodynamic drag forces from 
neighboring rowers and, thereby, prevents the rower to trespass the 
switching point significantly. In Fig. \ref{potential} we illustrate the 
reduced potential for $\alpha = 0.2$ and for two curvature values, 
$\epsilon =0$ and $-0.7$. The geometric switch variable is represented 
by solid ($ \sigma = 1$) and dashed ($ \sigma = -1$) lines, respectively.
With the choice of a positive $\alpha = 0.2$, the fast power stroke is 
performed at $ \sigma = 1$ when the rower moves in positive $y$ direction.

Note that Cosentino Lagomarsino \textit{et al.} \cite{lagomarsinoEtAl} define the 
power and recovery stroke by introducing different Stokes mobilities
in the forward ($ \sigma = 1$) and backward ($ \sigma = -1$) motion.
Such a choice does not directly affect the hydrodynamic-interaction terms
in Eq. (\ref{dynamiceq}) since the driving forces are the same. We tested
that this method does not effectively break the left-right symmetry of the
rower chain in order to create metachronal waves which run in one direction
along the whole chain. In contrast, by altering the strength of the 
driving force we are able to break the directional symmetry of the chain and
induce metachronal waves with a definite direction.



A single rower performs now the following oscillatory motion in a potential 
with negative curvature as illustrated in Fig.\ \ref{potential}. It starts 
off slowly from a switching point (e.g. at $y_m = -1$) and increases its 
velocity until it reaches the opposing switching point (e.g. at $y_m = 1$).
Here the switch variable $\sigma $ changes sign and the rower is 
dragged back into the opposite 
direction. Both, direction and magnitude of the velocity change 
discontinuously at the switching points. The non-zero curvature $\epsilon$ 
of the potential increases the period of a single rower relative to $\tau$,
the period when only a constant force $F_0$ acts.

\section{Synchronization of two rowers}
\label{sec_sync_2_rowers}

\begin{figure}
\centering
\subfloat{ \includegraphics[scale=0.9]{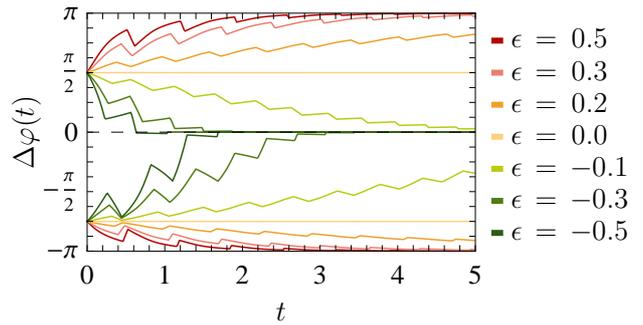} }
\caption{The synchronization of two rowers depends on the curvature 
$\epsilon $ of the driving-force potential. The time evolution of the
relative phase $\Delta \varphi = \varphi_2 - \varphi_1$ is shown for two 
intitial values, $\Delta \varphi = \pi/2 $ and $ -3\pi/4 $. Power and 
recovery stroke are symmetric, \textit{i.e.} $ \alpha = 0 $, the tilt angle $\beta = \pi/4$, and the distance 
of the rowers is $ c = 0.28 $.
}
\label{syncr_of_two}
\end{figure}

Figure\ \ref{syncr_of_two} summarizes our numerical studies on the dynamics 
of two rowers. The time evolution of the relative phase 
$\Delta \varphi = \varphi_2 - \varphi_1$ crucially depends on the curvature
$\epsilon$ of the driving-force potential. While positive $\epsilon$ leads
to antiphase synchronization regardless of the initial phase difference,
the rowers assume the same phase when $\epsilon$ is negative. For a 
constant force ($\epsilon=0$), $\Delta \varphi$ keeps its initial value
and does not change in time. We now develop an understanding for this
behavior.


For zero curvature $\epsilon$ of potential $V$ and identical power and 
re\-co\-ve\-ry stroke, $\alpha = 0$, the driving force on both particles
is constant and identical throughout one beating cycle. Then the mutual 
hydrodynamic interactions are always equal in absolute value and thus the
phase difference between both rowers does not change. The latter is even 
valid for the phase difference after one beating cycle if 
$\alpha \neq 0 $. Due to the different forces in the power and recovery
stroke, particles change their relative phase when they move in opposite
direction. However, during one beating cycle each particle performs both
the power and the recovery stroke. So, after one period the phase 
difference assumes its initial value and synchronization does not
occur. We stress again that $ \alpha $ is significant only
for breaking the left-right symmetry in a chain of rowers in conjunction 
with the tilt angle $\beta$ which then enables metachronal waves that travel
in one direction only.



A non-zero curvature in the driving-force po\-ten\-tial breaks the symmetry 
in the hydrodynamic interactions between two rowers and leads to 
synchronization as already demonstrated in Fig.\ \ref{syncr_of_two}.
Whilst Cosentino Lagomarsino \textit{et al.} restrict $ \epsilon $ to be positive, 
which leads to an antiphase synchronization \cite{lagomarsinoEtAl,kotarEtAl}, 
we also allow negative values, thereby obtaining the desired in-phase
synchronization. The significant part of the synchronization is accomplished 
when both rowers move in opposite directions. We first consider the case
$\epsilon < 0$ and assume that the preceding rower has just reached its
switching point. After activating the geometrical switch, the preceding
rower experiences a weaker driving force than the succeeding rower
before the switching point (see Fig.\ \ref{potential}). Therefore, the
flow field initiated by the succeeding rower is stronger and slows down 
the preceding rower more strongly than vice versa.
This leads to a decrease
of the phase difference $\Delta \varphi$ until the succeeding rower also
reaches the switching point. Then both rowers move in the same direction
and the phase difference $\Delta \varphi$ increases again. However, in 
total a decrease is realized during one cycle as illustrated in 
Fig.\ \ref{syncr_of_two}. For positive $\epsilon $, the preceding rower
slows down the succeeding rower more strongly and $\Delta \varphi$ 
increases on average during one cycle until it reaches the value $\pi$.
To conclude, in the time evolution of the phase difference 
$\Delta \varphi = 0$ is a stable and $\Delta \varphi = \pm \pi $ is an 
unstable fixpoint for $\epsilon < 0$, whereas for positive $\epsilon$
the stability of the fixpoints is reversed.

In a very attractive experimental realization of the two-particle system, 
Kotar \textit{et al.} use optical traps to create the driving-force 
potential with positive curvature $\epsilon >0$ \cite{kotarEtAl}. 
Each colloidal bead is confined in the harmonic regime of the potential 
well created by a laser beam. The potential well switches between two 
positions such that a bead is always driven towards the potential minimum.
The ``geometric switch'' is then initiated before the bead reaches the
minimum. In agreement with our results, the two beads synchronize in 
antiphase. The trap potential created by an optical tweezer has a regime
further away from the minimum, where the curvature is negative
\cite{opticaltweezers}. We propose to use this regime to realize 
in-phase synchronization experimentally.


\section{Synchronization in a chain of rowers} \label{sec_sync_chain}

\subsection{Complex order parameter} \label{subsec_cplx_ord-parameter}

The collective dynamics of a chain of several hundred rowers strongly 
depends on the parameters introduced above that determine the 
driving-force potential and the geometry of the rower chain.
In order to classify the collective dynamics and, in particular, to 
identify metachronal waves, we introduce a complex order parameter
and study its longterm behavior.


For an array of $N$ oscillators, we can introduce $N-1$ phase differences 
$ \{ \Delta\varphi_{1}, \dots, \Delta\varphi_{N-1} \} $ and their phasors 
$z_{n} = e^{ i \Delta \varphi_{n}} \in \mathbb{C} $, where 
$\Delta\varphi_{n} = \varphi_{n+1} - \varphi_{n} $. 
We define our complex order parameter as the mean of the phasors,
\begin{equation}
Z = A e^{ i \varPhi } := \frac{1}{N-1} \sum\limits_{n = 1}^{N-1} z_{n} = 
     \frac{1}{N-1} \sum\limits_{n = 1}^{N-1} e^{ i \Delta \varphi_{n} } 
\enspace .
\end{equation}
The magnitude $A$ and polar angle $\varPhi$ lie in the respective ranges
$A \in [ 0, 1 ] $ and $ \varPhi \in [ - \pi, \pi ) $. We add a note here.  
In order to determine if oscillators fully synchronize to equal phases, 
usually the order paramter averages over the phasors $e^{ i \varphi_{n}} $ 
for phases $\varphi_{n}$ \cite{syncr_Definitions,syncrRosenblumEtAl}. 
However, this definition is not appropriate for identifying metachronal
waves.


For a large number $N$ of rowers with random phases, the phasors 
$z_{n}$ are uniformly distributed over the unit circle in the complex plane.
Their mean, the complex order parameter $Z$, is close to 0, 
\textit{i.e.}, $A \approx 0 $ with arbitrary $ \varPhi $. If, on the 
other hand, the system displays a stable metachronism, where pairs of 
neighboring oscillators are phase locked with the same phase difference 
\cite{syncr_Definitions}, then $ z_{n} = z_{m} $ for all phasors $ m, n \in 
\{ 1, \dots , N - 1 \} $. In this case, $ Z $ is stationary. It
lies on the unit circle in the complex plane ($A = 1$) and $ \varPhi $ is 
equal to the phase difference of each pair of oscillators.


In the following, we will also observe states where the long-time limit of
$Z$ fluctuates around a constant value. We therefore introduce the 
average values
\begin{equation}
\bar{A} = \frac{1}{\Delta t} \int_{T - \Delta t}^{T} A(t) dt \qquad
\bar{\varPhi} = \frac{1}{\Delta t} \int_{T - \Delta t}^{T} \varPhi(t) dt 
\enspace,
\label{average_rule}
\end{equation}
where $T$ is the final time of the simulation run and $ \Delta t $ an
appropriately chosen time interval. In addition, we also
quantify fluctuations of the order parameter by calculating the
standard deviations $\sigma_{A}$ and $\sigma_{\varPhi}$ of the
time averaged values.

\subsection{Synchronization in an unbounded fluid}
\label{subsec_unbound_fluid}

\begin{figure}
\centering
\subfloat[]{
\includegraphics[scale=0.9]{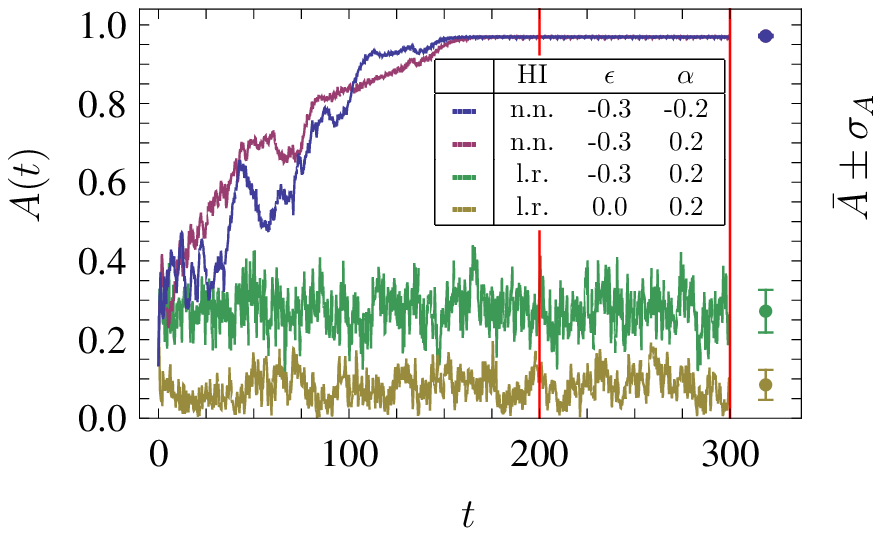}
}\\
\subfloat[]{
\includegraphics[scale=0.9]{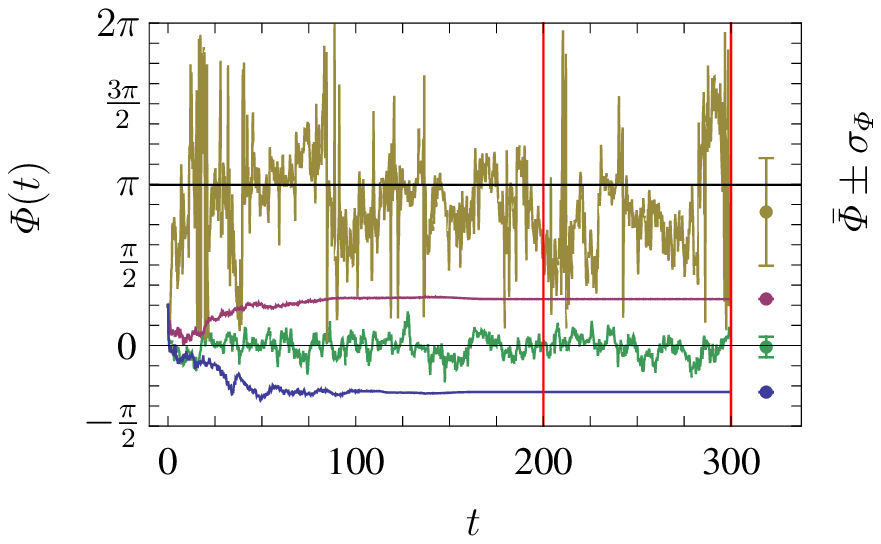}
}
\caption{Absolute value (a) and polar angle (b) of the order parameter 
$Z(t)$ as a function of time for a chain of 200 rowers for different
parameter sets. The global parameters are $c=0.28$ and $\beta = \pi / 4$ 
and for all graphs the rowers start with the same random distribution of 
phases. The vertical red lines ({\color{red}\legendLn}) mark the 
time intervall $\Delta t = 100$ for which we calculate the time averages 
$\bar{A} $ and $\bar{\varPhi}$. They are shown as dots on the right-hand
side of the corresponding diagrams together with errorbars indicating
the respective standard deviations $\sigma_{A}$ and $ \sigma_{\varPhi}$.
Golden line ({\color{g_gold}\legendLn }): $ \alpha = 0.2$ , $\epsilon = 0$, 
and long-range hydrodynamic interactions  (l.r. HI)
are used. Note for this parameter set we have added $2 \pi$ to all 
negative values of $\varPhi (t)$ to better represent its time dependence.
Green line ({\color{g_green}\legendLn }): $ \alpha = 0.2$ and 
$\epsilon = - 0.3$ with l.r. HI.
Magenta line ({\color{g_red}\legendLn}): $ \alpha = 0.2$ 
and $ \epsilon = -0.3 $ but hydrodynamic interactions are artificially
restricted to nearest neighbors (n.n. HI). A metachronal wave occurs 
that travels in a definite direction. Blue line 
({\color{g_blue}\legendLn}): same situation but $\alpha = -0.2 $. 
The mean phase $\bar{\varPhi}$ is reversed and the metachronal wave travels 
in the opposite direction.
}
\label{absValNangle_over_time_D2}
\end{figure}

\begin{figure}
\subfloat[]{
\includegraphics[scale=0.88]{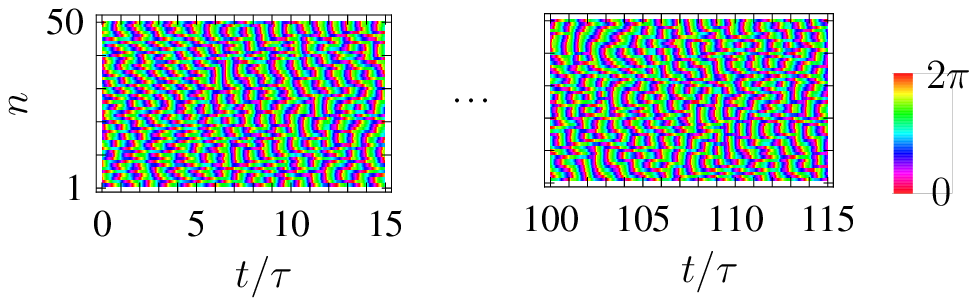}
}\\
\subfloat[]{
\includegraphics[scale=0.88]{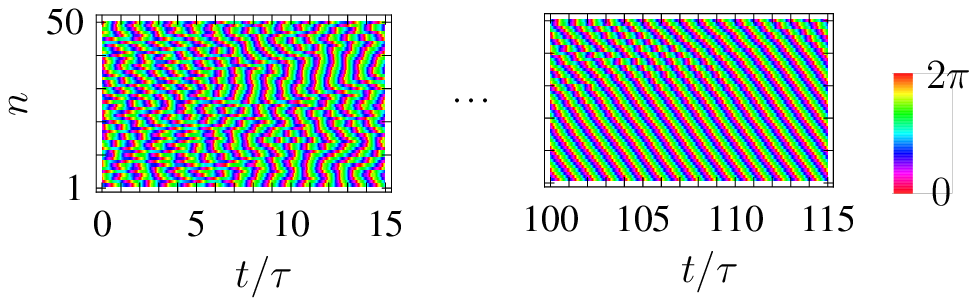}
}
\caption{Color coded phases of the the first 50 rowers in a chain of 200
rowers as a function of time $t$. The global parameters are $\alpha = 0.2$, 
$c=0.28$, and $\beta = \pi/ 4$. The simulations start with $\epsilon = 0$ and 
at $t = 5$ the curvature changes to $ \epsilon = -0.3 $ in order to visualize 
the onset of synchronization. (a) For long-range hydrodynamic interactions
transient in-phase synchronization occurs locally. (b) For nearest-neighbor
interactions metachronal waves are visible.}
\label{fig_phase_colorPlot}
\end{figure}


In Figs.\ \ref{absValNangle_over_time_D2}(a) and \ref{absValNangle_over_time_D2}(b) 
we  illustrate the temporal evolution of the order parameter $Z(t)$ for an 
open chain of 200 rowers in different situations.
For $ \epsilon = 0 $ and $\alpha = 0.2$ [golden graph 
({\color{g_gold}\legendLn })], magnitude $A(t)$ and, in particular,
$\varPhi(t)$ strongly fluctuate in time. Their respective mean values
are $\bar{A} \ll 1$ and $\bar{\varPhi} \approx \pi$. Apart from a tendency 
of neighboring rowers to beat in antiphase, a noticable structure formation
does not occur.


We now introduce a negative curvature $\epsilon = -0.3 $ in the 
driving-force potential. The green graphs ({\color{g_green}\legendLn })
in Figs.\ 
\ref{absValNangle_over_time_D2}(a) and \ref{absValNangle_over_time_D2}(b)
still show a strongly fluctuating order parameter. In contrast to 
$ \epsilon = 0 $ though, we observe a significant increase of the 
time-averaged magnitude $\bar{A}$. Together with $\bar{\varPhi} \approx 0$ 
this indicates that some of the rowers synchronize in phase.
A further understanding is achieved from Fig. \ref{fig_phase_colorPlot}(a),
where we plot the color-coded phases over time for the first 50 rowers.
At $t=0$ we start with curvature $\epsilon=0$ and switch to 
$\epsilon = -0.3 $ at $t= 5$. Then chain segments of approximately 8 rowers
appear that are fully synchronized in phase as expected for negative
curvature. However, after some time these clusters break apart and 
form somewhere else. The synchronization is only transient since the 
rowers in the center of a cluster have reduced friction coefficients 
compared to rowers at the edges due to long-range hydrodynamic interactions. 
Hence, rowers in the center move faster and destabilize the cluster.



The reported instability is caused by the long-range nature of
hydrodynamic interactions. Following Ref. \cite{niedermayerEtAl}, we 
artificially make them short-ranged by including only the nearest neighbors
in the sum of the dynamic equations (\ref{dynamiceq}). The range of
hydrodynamic interactions can be controlled close to a surface, as we
demonstrate in the next section. Now metachronal waves appear that
travel from one end of the chain to the other with a wavelength of 
approximately eight rowers. Fig. \ref{fig_phase_colorPlot}(b)
demonstrates impressively how the initially uncorrelated rowers evolve
into a metachronal wave due to phase-locking. The process is completed
for the whole chain after $t=175$. This is also visible in Fig. 
\ref{absValNangle_over_time_D2}(a) [magenta line ({\color{g_red}\legendLn})]
when $ A(t) \approx 1 $ is achieved. The phase difference 
[Fig. \ref{absValNangle_over_time_D2}(b)] converges to approximately
$\bar{\varPhi} \approx \pi / 4 $ which gives the corresponding wavelength
of 8 rower distances. The very small standard deviations
of $\bar{A}$ and $\bar{\varPhi}$ are a further measure that a metachronal 
wave has been formed nearly perfectly. Finally, reversing the
sign of $\alpha$ [blue curves ({\color{g_blue}\legendLn}) in 
Figs. \ref{absValNangle_over_time_D2}(a) 
and \ref{absValNangle_over_time_D2}(b)], which means exchanging the 
fast power and slow recovery strokes, also reverses the sign of 
$\bar{\varPhi}$ and the metachronal wave travels in the opposite 
direction.


To break the left-right symmetry of the rower chain, parameter values
$\alpha \ne 0$ and $\beta \ne \pi/2$ are needed but also the 
distance $c$ of the rowers has to be sufficiently small for creating 
metachronal waves travelling in one direction. For $\alpha= 0$ or 
$\beta = \pi/2$ or by simply choosing $c>1$, we observe metachronal 
waves running from the center of the chain into both directions. 
This is reminiscent to studies of Niedermayer \textit{et al.} 
\cite{niedermayerEtAl}, where beads move on circular trajectories.



Positive curvature $ \epsilon > 0 $ of the driving-force potential 
reproduces the results of Cosentino Lagomarsino \textit{et al.} 
\cite{lagomarsinoEtAl}. The tendency of neighboring rowers to synchronize 
in antiphase leads to a metachronism with shorter wavelengths of 
approximatly 4 rowers. Compared to the case with $ \epsilon = -0.3$,
the system needs about ten times longer to reach the stationary state.


	
\begin{figure}
\centering
\subfloat{ \includegraphics[scale=0.9]{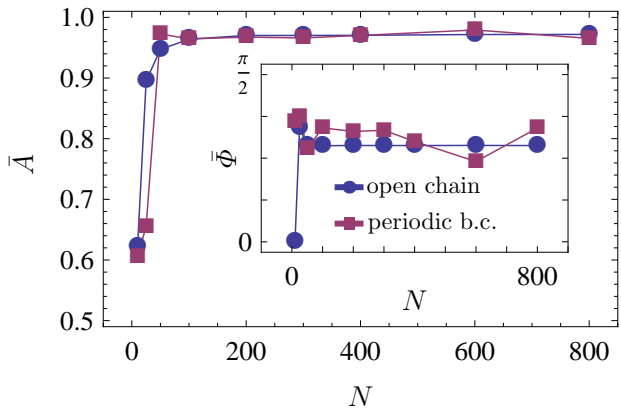} }
\caption{Averaged magnitude $ \bar{A} $ and polar angle $\bar{\varPhi}$
(inset) of the order parameter as a function of the number $N$ of rowers in
open (circles) and closed (squares) chains.
}
\label{fig_open-vs-closed_chain}
\end{figure}

We shortly discuss how the chain length of open chains and also periodic 
boundary conditions influence the occurence of metachronal waves. We
concentrate on the same parameters as before ($\epsilon = -0.3$, 
$\beta =\pi/4$, and $ \alpha = \pm 0.2 $) and use hydrodynamic 
interactions restricted to nearest neighbors. This allows us to realize 
the periodic boundary conditions by a closed chain where the first and 
last rower interact with each other. Figure \ref{fig_open-vs-closed_chain} 
compares $\bar{A}$ and $\bar{\varPhi}$ (inset) for open (circles)
and closed (squares) chains for different chain lengths.
For chain lengths above 50 rowers open and closed chains develop
metachronal waves with nearly the same $ \bar{A} $ and $\bar{\varPhi}$.
So finite size effects become negligible. At shorter chain lengths the
dynamics of open and closed chains differ strongly. Whereas for 
$N = 10$ (first data point) the same magnitude $\bar{A} \approx 0.6$ 
indicates some order but certainly no metachronism, the polar angles 
reveal different behavior in the two cases.
For open chains $\bar{\varPhi} = 0$, which means that central rowers 
synchronize in phase whereas the outer rowers lag behind due to the 
larger friction they experience. On the other hand, in closed
chains, where all rowers have an identical environment, a non-zero mean polar 
angle $\bar{\varPhi}$ is observed. A closer inspection of the individual
rower phases shows that pairs of rowers almost fully synchronize in 
phase whereas they leave a larger phase gap to the next pair. We discuss 
these points so detailed here because in an experimental realization 
finite size effects of the chain may become important.

\subsection{Synchronization near an infinitely extended wall}
\label{subsec_near_wall}

\begin{figure}
\centering
\subfloat{ \includegraphics[scale=1.25]{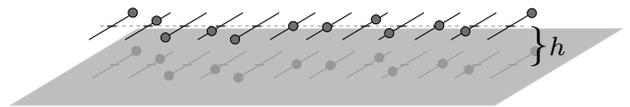} }
\caption{A chain of rowers a distance $h$ above an infinitely 
extended wall.}
\label{chain_over_wall}
\end{figure}

\begin{figure}
\centering
\subfloat{ \includegraphics[scale=0.9]{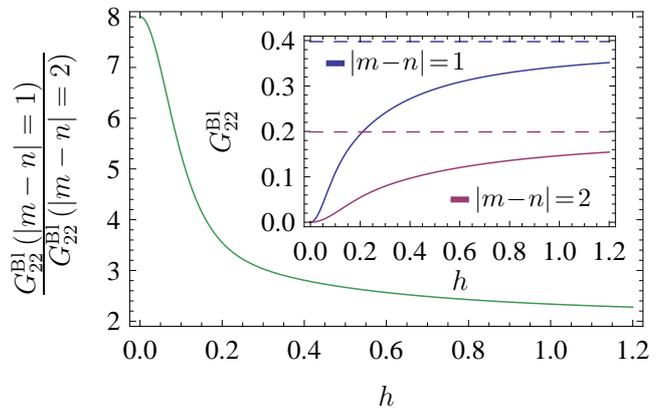} }
\caption{Inset: The relevant component $G_{22}^{\text{Bl}}$ of the Blake 
tensor as a function of $h$ for, both, nearest-neighbor ($|m-n|=1$) and 
next-nearest neighbor ($|m-n|=2$) rowers. The distance of the 
nearest-neighbor and next-nearest neighbor rowers are $c=0.28$ and 
$2c=0.56$, respectively. The dashed lines indicate the respective values 
of the Oseen tensor, \textit{i.e.}, for height $h \rightarrow \infty$. 
Large graph: ratio of both components, 
$G_{22}^{\text{Bl}}(|m-n|=1)/ G_{22}^{\text{Bl}}(|m-n|=2)$.
}
\label{HI_range}
\end{figure}

So far we artificially restricted hydrodynamic interactions to nearest
neighbors in order to make them short-ranged. In reality, the range of 
hydrodynamic interactions can be tuned by placing the chain of rowers
a distance $h$ above an infinitely extended wall. As sketched in
Fig.\ \ref{chain_over_wall}, the rowers beat parallel to this wall.
In order to describe the cross mobilities of point particles close to
a wall, the Oseen tensor has to be replaced by the Blake tensor
\cite{blake_singularities_1974_etAl}. The latter converges to the
Oseen tensor for $ h \rightarrow \infty $. So, by continuously changing
the height $h$, one can control the range and strength of hydrodynamic 
interactions. They decay with $1/r$ at large distances $h$ from the wall, 
thus being of long range. In the vicinity of the wall, they decay with 
$1/r^3$ and justify the approximation of nearest-neighbor interactions
used in Ref. \cite{niedermayerEtAl}. In Fig.\ \ref{HI_range} in the inset 
we plot the relevant component $G_{22}^{\text{Bl}}$ of the Blake tensor 
as a function of $h$ for, both, nearest-neighbor and next-nearest-neighbor 
rowers. The dashed lines indicate the respective values of the Oseen 
tensor that are reached in the limit $ h \rightarrow \infty $. On the other 
hand, for $h \rightarrow 0$ the relative strength of the nearest-neighbor 
component $G_{22}^{\text{Bl}}$ compared to the next-nearest-neighbor 
component increases as the large graph in Fig.\ \ref{HI_range} 
demonstrates. So, we expect to observe with decreasing $h$
a transition from transient synchronization to the formation of 
metachronal waves.

\begin{figure}
\centering
\subfloat[]{
\includegraphics[scale=0.9]{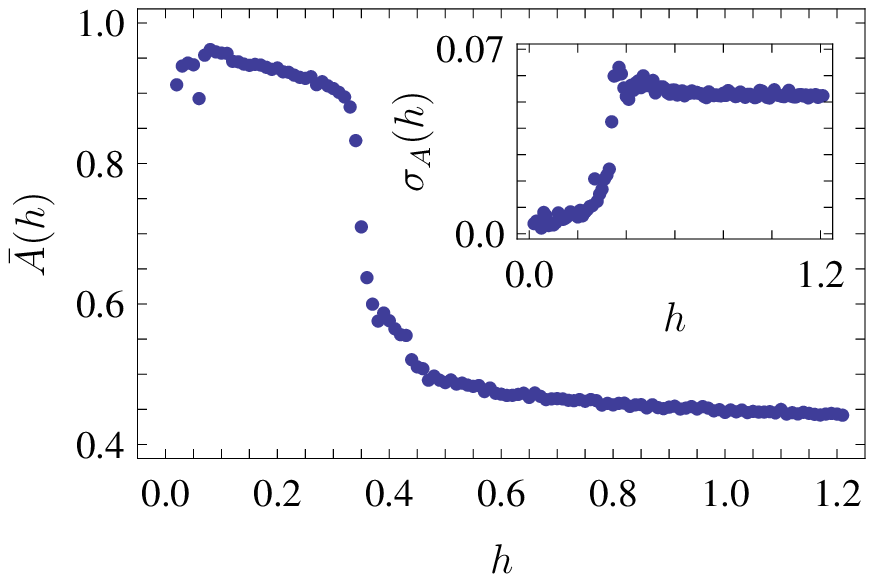}
}
\\
\centering
\subfloat[]{
\includegraphics[scale=0.9]{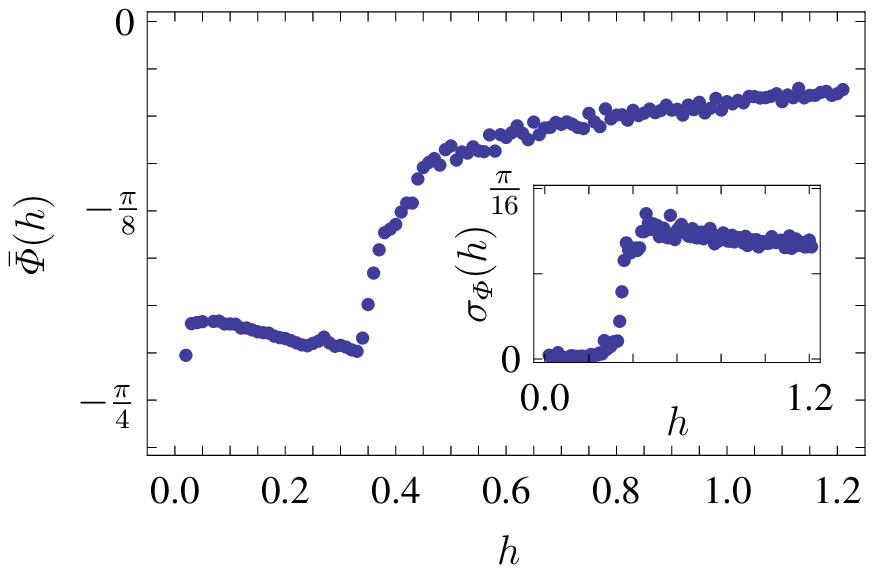}
}
\caption{Average magnitude (a) and polar angle (b) of the order parameter
$Z$ for a rower chain situated at a height $h$ above a bounding wall.
The insets show the respective standard deviations. The global
parameters of the open chain with $ N = 200 $ rowers are $\beta = \pi/4$, 
$ c=0.28 $,  $ \alpha = -0.2 $, and  $\epsilon = -0.4 $. For $h > 0.15$, 
the total simulation time was $T=3750$ and the time window for 
determining the averages was chosen sufficiently large, $ \Delta t = 2500$,  
in order to minimize fluctuations of $\bar{A}$ and $\bar{\varPhi}$. For 
$h \le 0.15$, the simulation time was extended to $T=6250$ to ensure that 
a stationary metachronal state was reached and $\Delta t = 125$.
}
\label{fig_ordParaAvr_over_height}
\end{figure}

\begin{figure}
\subfloat{ \includegraphics[scale=1]{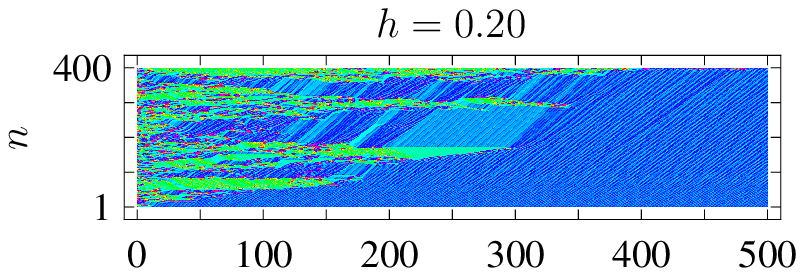} }
\vspace{3mm} \\
\subfloat{ \includegraphics[scale=1]{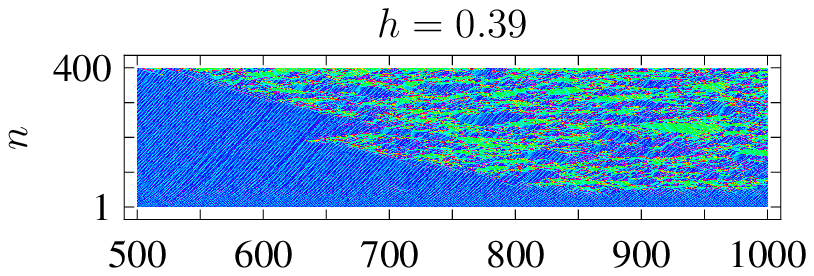} }
\vspace{3mm} \\
\subfloat{ \includegraphics[scale=1]{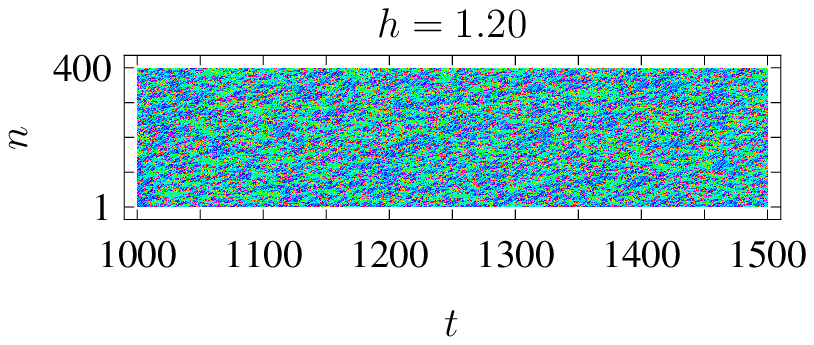} }
\vspace{3mm} \\
\hspace*{28mm} \subfloat{ \includegraphics[scale=1]{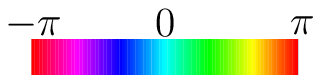} }
\caption{Color-coded phase differences of neighboring rowers as a function
of time in a chain of 400 rowers. The simulation starts at a height of $ h = 0.20 $,
which is then increased to $h = 0.39$ and $h = 1.20$ at times $t=500$ and
$t=1000$, respectively. At $ h = 0.20 $ a metachonal wave develops 
and finally stretches over the entire chain. At $ h = 0.39 $ the metachronism 
starts to break apart and transient synchronization with larger patches of
correlated rowers are visible. At  $h = 1.20 $ the patches become
smaller.
%
%
}
\label{colorPlot_phaDiff}
\end{figure}


This is demonstrated in Figs.\ \ref{fig_ordParaAvr_over_height}(a) and 
\ref{fig_ordParaAvr_over_height}(b), where we plot the time-averaged 
magnitude $\bar{A}$ and polar angle $\bar{\varPhi}$ of the order parameter 
$Z$ as a function of the height $h$. The insets depict the corresponding 
standard deviations. 
At hights $h$ smaller than 0.35, $ \bar{A}  > 0.9$ clearly indicates
metachronism with an absolute phase difference between rowers smaller
than $\pi/4$. It only varies slightly with $h$ so that the wavelengths of 
the metachronal waves correspond to 9 to 10 rowers. Also the very small 
standard deviations are typical for
metachronism. Then a relatively sharp transition occurs at $h = 0.35$ 
into a state of transient synchronization. The amplitude $ \bar{A}$
decreases by a factor of two and the negative $\bar{\varPhi}$ grows 
towards zero. In particular, the destruction of metachronism is indicated
by a strong increase in the standard deviations showing that amplitude $A$ 
and polar angle $\varPhi$ of the order parameter fluctuate strongly.
In Fig.\ \ref{colorPlot_phaDiff} the transition is also visible. We show 
the color-coded phase differences between neighboring rowers
as a function of time. At $t =0$ we start at a hight $h=0.20$ and then
during the simulation switch to $h=0.39$ at $t =500$ and to $h=1.20$
at $t = 1000$. Whereas the nearly uniform blue region at $h=0.20$ 
shows the existence of metachronal waves, they dissolve with time 
at $h=0.39$ and transient patches of same color indicate some
synchronization. At $h=1.20$ the patches become smaller 
corresponding to a further decrease in $ \bar{A} $ and the results are similar
to our findings in the unbounded fluid [compare green graphs in
Figs.\ \ref{absValNangle_over_time_D2}(a) and 
\ref{absValNangle_over_time_D2}(b)].

Our results show that me\-ta\-chro\-nal waves in a chain of in-phase
synchronizing rowers can only develop in the vicinity of a surface,
where the range of hydrodynamic interactions is strongly reduced.
Long-range hydrodynamic interactions destabilize the waves since 
they lead to the formation of transient clusters synchronized in-phase.

%


\section{Conclusions} \label{sec_conclusion}

In order to shed light on the essential features for obtaining metachronal
waves, we implemented a chain of oscillators or rowers that are driven
by an external non-linear force derived from a driving-force potential.
Two rowers synchronize in antiphase when the curvature of the force 
potential is positive or in phase when the curvature is negative.
We concentrated on the latter case and introduced a suitable order parameter
for identifying metachronal waves. Our investigations show that metachronal
waves exist only when the range of hydrodynamic interactions is restricted
either artificially or by the presence of a bounding surface as in any
relevant biological system. The negative curvature in the driving force
potential leads to me\-ta\-chro\-nal waves with wavelengths of 7-10 rower
distances. Curiously, in paramecium a wavelength including eight cilias
is observevd \cite{machemerParamecium}. In comparison, positive 
curvature that promotes antiphase synchronization in a rower pair only
gives shorter wavelengths of around four rowers \cite{lagomarsinoEtAl}.
In order to obtain waves that travel either in one or the other direction
along the chain, we break its left-right symmetry by tilting the rower
segment with respect to the normal of the chain and distinguish between
a fast power and slow recovery stroke. These measures are only effective
when the rowers are sufficiently close to each other ($c<1$).

Me\-ta\-chro\-nal waves and their defining properties, such as the 
wavelength and the propagational direction relative to the direction of 
the power stroke, are an essentially two-di\-men\-sio\-nal phenomenon
as Machemer pointed out \cite{machemerParamecium}. 
Therefore, it will be important to extend current research towards models
defined on two-dimensional lattices. This should give further insight
into the fascinating properties of metachronal waves. Theoretical work 
in this direction was carried out on microfluidic rotors 
\cite{Grzybowski:2000,Lenz:2003,golestanian2010EtAl}. However,
amongst the great variety of beautiful patterns evolving, in particular,
in Ref. \cite{golestanian2010EtAl}, metachronal waves as seen on 
ciliated surfaces cannot be observed. We have recently started to
look at two-dimensional regular arrays of mainly stiff rods attached to a 
surface.  We let them perform strokes where they move on a cone tilted 
with respect to the normal of the surface \cite{Downton10}. When we divide 
the beating cycle into a fast power and a slow recovery stroke, the latter 
acting when the rod is close to the surface, metachronal waves emerge.
For a planar stroke see Ref.\ \cite{Elgeti06}.
We will continue working on our system and also think about closed 
curved surfaces where topological constraints arise.

\appendix

\section{Mobilities for a fluid bounded by a plane wall} \label{app.wall}

When the fluid is bounded by an infinitely extended wall in the $xy$\ plane,
the mobilities for point-like particles assume the following expressions. 
The self-mobility depends on the distance $h$ from the bounding wall 
\cite{DufresneEtAl}:
\begin{equation}
\vc{M}_{mm}(\vc{r}_{m}) = \mu_{0}  \left[ \mathbbm{1} - 
  \frac{9}{16} \frac{a}{h} \begin{pmatrix}
			        1 & 0 & 0 \\
			        0 & 1 & 0 \\
			        0 & 0 & 2 \\
                           \end{pmatrix} \right] \, ,
\end{equation}
and the Blake tensor $ \vc{G}^{\text{Bl}} (\vc{r})$ provides the cross 
mobilities
\begin{eqnarray}
\vc{M}_{mn}(\vc{r}_{m},\vc{r}_{n}) & = 
& \vc{G}^{\text{Bl}} (\vc{r}_{m},\vc{r}_{n}) \nonumber\\ 
& = & \vc{G}^{ \text{Os} } (\vc{r}_{m} - \vc{r}_{n} ) - \vc{G}^{ \text{Os} } ( \vc{r}_{m} - \bar{ \vc{r} }_{n} ) \\
& & + 2 h^{2} \vc{G}^{\text{So}} ( \vc{r}_{m} - \bar{\vc{r}}_{n} ) - 2 h \vc{G}^{\text{St}} ( \vc{r}_{m} - \bar{\vc{r}}_{n} ) \, ,
\nonumber
\label{eq_Blake-Tensor}
\end{eqnarray}
where $ \vc{r}_{n} = (x_{n}, y_{n}, h) $ is the location of the point force
and $ \bar{\vc{r}}_{n} = (x_{n}, y_{n}, -h) $ is its image point 
\cite{vilfanEtAl,blake_singularities_1974_etAl}. Using the Kronecker symbol 
$\delta_{\lambda\nu} $, the components of the so-called source doublet 
$\vc{G}^{ \text{So}} $ and Stokes doublet $ \vc{G}^{ \text{St} } $ read,
respectively,
\begin{align}
G_{ \lambda \nu}^{ \text{So} } ( \vc{x} ) & =
	\frac{1}{8 \pi \eta} ( 1 - 2 \delta_{ \nu 3 } ) \frac{ \partial }{ \partial x_{ \nu } } \left( \frac{ x_{ \lambda } }{ \lvert \vc{x} \rvert^{3} } \right) \\
G_{ \lambda \nu}^{ \text{St} } ( \vc{x} ) & =
   ( 1 - 2 \delta_{ \nu 3 } ) \frac{ \partial }{ \partial x_{ \nu } }
   G_{ \lambda 3}^{ \text{Os} } ( \vc{x} ) \, .
\end{align}


%

\end{document}